\documentclass[pre,showpacs,amsmath,amssymb]{revtex4}  

\usepackage{graphicx}
\usepackage[colorlinks=false,linkcolor=blue]{hyperref}  
\begin{document}                
\preprint{PRE/RT-CBS}
\def\convo{\operatorname*{\otimes}}
\newcommand{\T}{\Theta}

\title{The relation between Time Reversal focusing and Coherent Backscattering in multiple scattering media: a diagrammatic approach }
\author{Julien de Rosny}
\email{julien.derosny@espci.fr}
\author{Arnaud Tourin}
\author{Arnaud Derode}

\affiliation{Laboratoire Ondes et Acoustique, CNRS/ESPCI/Universit\'e Paris VII, UMR
7587, 10 rue Vauquelin, 75005 Paris, France}
\author{Bart van Tiggelen}
\affiliation{Laboratoire de Physique et Mod\'elisation des Milieux Condens\'es, CNRS/Maison de Magist\`eres, Universit\'e Joseph Fourier B.P. 166, 38042 Grenoble Cedex 9, France}

\author{Mathias Fink}
\affiliation{Laboratoire Ondes et Acoustique, UMR
7587, CNRS/ESPCI/Universit\'e Paris VII, 10 rue Vauquelin, 75005 Paris, France}

\date{\today}
\begin{abstract}                
In this paper, we revisit one-channel time reversal (TR) experiments through multiple scattering media in the framework of the self-consistent multiple scattering theory. The hyper resolution and the self-averaging property are retrieved. The developed formalism leads to a deeper understanding of the role of the ladder and most-crossed diagrams in a TR experiment, and also establishes the link between TR and Coherent Backscattering (CBS). Especially, we show that when the initial source and the time reversal point are at the same location, the time-reversed amplitude is twice higher.  Surprisingly, this enhancement is due to the ladder diagrams and not to the most-crossed ones, contrary to CBS. These theoretical predictions are confirmed by experimental results. The experiments are performed with ultrasonic waves propagating through a random collection of parallel steel rods. 
\end{abstract}

\maketitle
\section{Introduction}               
A few years ago, we have carried out the first experimental demonstration of the reversibility of acoustic waves propagating  through a 2-dimensional random collection of scatterers, the ultrasonic equivalent of a Lorentz gas for particles\cite{Derode95}. Typically, an ultrasonic source sends a broadband pulse into a medium consisting of thousands of parallel steel rods randomly distributed and immersed in water. Since the mean free path is much less than the sample thickness, strong multiple scattering occurs. The multiply scattered signals transmitted through the medium are recorded on a 128-transducer array, digitized and time-reversed. A part of the signal, the so-called ``time reversal window'', is transmitted back into the medium. The time reversed wave is found to converge back to its source. Two aspects of this problem have been studied~: the signal recreated at the source location (time compression) and the spatial focusing of the time-reversed wave around the source location. Dowling and Jackson\cite{Dowling92} have been pioneers in this field: three years before the first experimental observation, they predicted the peculiar property of super-resolution of time reversal in random media. But their approach was restricted to narrow-band signals and time-compression was thus not expected. Later we developed a phenomenological statistical model to describe both spatial and temporal focusing of a broadband pulse\cite{Derode99,Derode00,Derode01}. Blomgren et al.\cite{Blomgren02} applied the Green's function formalism to analyze the spatio-temporal focusing obtained by time reversal in a random medium with weak celerity fluctuations.  However, their model was based on the parabolic (or paraxial) approximation\cite{Tappert77} which assumes only weak scattering angles and thus ignores backscattering. In this latter paper, the reason why the focusing is observed for one realization of disorder is referred to as ``pulse stabilization''. It comes from the ``self-averaging'' behavior of the time reversal wave due to the broadband character of the initial pulse. 
  
In the present paper we revisit time reversal experiments in the framework of the multiple scattering theory. This theory describes statistical moments of the scattered field in transmission as well as in backscattering. It has been extensively applied in several fields of physics (optics, electronics, acoustics, ...). It is a general theoretical framework that successfully describes various multiple scattering phenomena: diffusive field fluctuations and their application to Diffusive Wave Spectroscopy (DWS)\cite{Maret87,Pine88}, weak and strong localization\cite{Wiersma97,Tiggelen99}, short and long range intensity correlations ($C_1$, $C_2$, $C_3$)\cite{Berkovits94}, etc... It was only very recently that arguments based on multiple scattering theory have been proposed in order to explain some results of TR in random media\cite{Tiggelen03}. For simplicity, we focus in this paper on time reversal performed with one transducer. The average time-reversed field and its variance are worked out. 
Comparing the mean value of the TR field to its standard deviation leads us to understand the hyper-resolution as well as the self-averaging property of TR in random media. In addition to this, one of the most striking results of this paper is the following: the time reversal amplitude is predicted to be twice larger when the time reversal element and the initial source positions are the same. This effect is closely related to the well-known Coherent Backscattering Effect (CBS)\cite{VanAlbada85,Wolf85}. Since the TR element is a receiver in the first step and a source in the second step of the TR process, the enhancement in TR focusing is due to the Ladder diagrams and not to the most-crossed ones, contrary to CBS! 

We first present some experimental results in order to introduce the questions that we address in this paper before presenting the theory in detail.  The theoretical developments are based on the diagrammatic theory of multiple scattering. We apply it to determine the average amplitude and the variance of the TR field in both transmission and backscattering. In particular, we show that time reversal focusing depends on the so-called Vertex,  which is the four-entry correlation function of the random wave-field. The study is limited to media whose Thouless number is much larger than one. In other words, the crossing paths inside the multiple scattering medium are neglected. The entire theory is explicitly developed in the time domain. We show that TR can be completely analyzed in terms of most-crossed and ladder diagrams. This general approach facilitates the study of two fundamental aspects of one-channel TR in multiple scattering media: hyper-resolution and self-averaging. We end the theoretical part by establishing a formal link between TR focusing and coherent backscattering.
Finally, we show experimental results that confirm the theoretical predictions.
\section{A first experiment}
\label{sec:preliminary}
\begin{figure}
\begin{center}
\includegraphics[width=7cm]{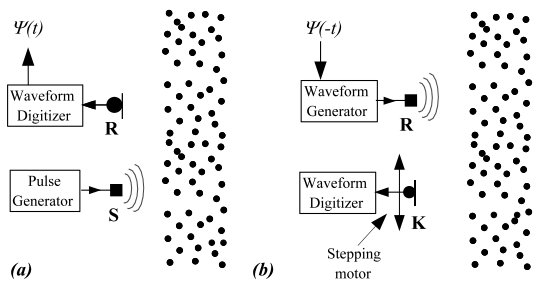}
\caption{\label{fig:setup}The two steps of a one-channel TR experiment through a multiple scattering medium in a backscattering configuration. (a) forward propagation step. (b) backward propagation step.}  
\end{center}
\end{figure}

\begin{figure}
        \begin{center}
        \includegraphics{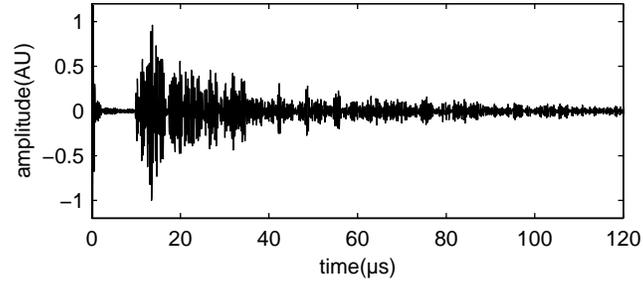}
        \end{center}
        \caption{\label{fig:impulse_response} A typical time-evolution for the pressure field recorded at point $\mathbf{R}$ after the emission of a short $1 \mu s$ pulse (at a central frequency of 3.5 MHz) at point $\mathbf{S}$.}
\end{figure}

\begin{figure}
        \begin{center}
        \includegraphics{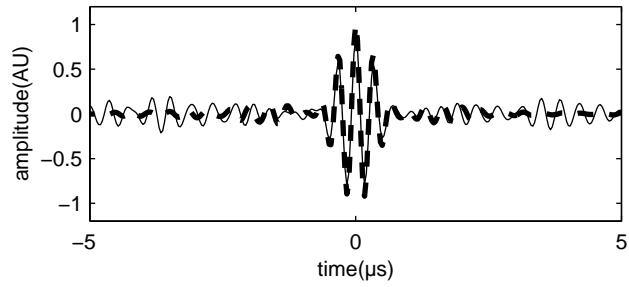}
        \caption{\label{fig:RT_time_foc} The continuous line corresponds to the time compression recorded at the initial source position ($\mathbf{K} = \mathbf{S}$) for one realization of disorder. The dotted line is the average value over 100 realizations of disorder. The signals are normalized by the maximum of the average signal which occurs at $t=0$.} 
        \end{center}
\end{figure}

\begin{figure}
        \begin{center}
        \includegraphics{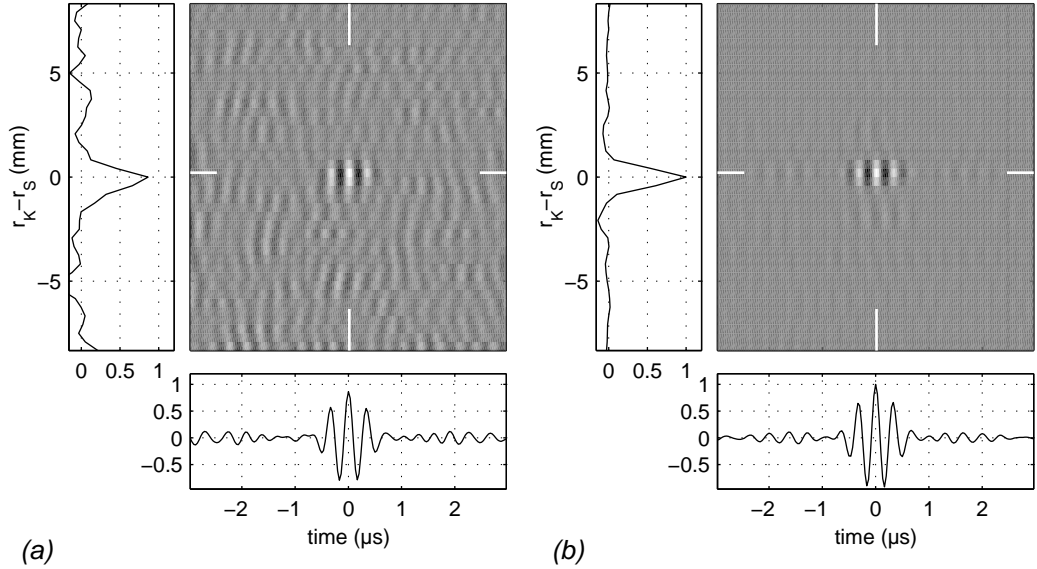}
        \caption{\label{fig:RT_time_foc2}Spatio-temporal focusing on a single realization of disorder (a) and averaged over 100 realizations (b). The vertical axis corresponds to the distance between the recording point ($\mathbf{K}$) and the initial source position ($\mathbf{S}$). The signals are normalized by the value of the averaged signal at $t=0$ and $\mathbf{K}=\mathbf{S}$}
\end{center}
\end{figure}

An acoustic wavefield can be time-reversed using a Time Reversal Mirror (TRM). A TRM is usually made of 128 independent time reversal channels. Each channel consists of a piezoelectric transducer plugged to a digital emitting/recording electronics. In this article, we only deal with one-channel TR experiments performed through multiple scattering media. Indeed, the physics in such a configuration is already sufficiently new, rich and complex to devote the complete study to this topic. 
The schematic view of the one-channel TR setup is presented in Fig.~\ref{fig:setup}.
The experiments have been carried out in a 2D-like multiple scattering medium made of thousands of steel rods. The sample is $35\:mm$ thick and $300\:mm$ large. There are 18.75 rods per square centimeter and the diameter of each rod is $0.8~mm$. The elastic mean free path is about $4 mm$\cite{Tourin00}. The time reversal experiment is divided in two steps. First, a small emitter (here a piezoelectric transducer) located at point $\mathbf{S}$ generates a short ultrasonic pulse at a central frequency of 3.5 MHz ($0.43mm$ wavelength) with a $100$-percent bandwidth toward the multiple scattering medium. Another piezoelectric transducer acting as a small microphone records the time-dependence of the backscattered wave field, $\Psi(t)$, at point $\mathbf{R}$. A typical backscattered signal is plotted in Fig.~\ref{fig:impulse_response}.
Since the sample thickness is much larger than the mean free path, the huge time spreading (more than $100\mu s$) of the initial pulse is due to multiple scattering. Secondly, the recorded signal is time-reversed and sent back into the medium by the transducer at point $\mathbf{R}$ now acting as a small loudspeaker. Finally a small transducer plugged to a waveform digitizer records the time-dependence of the back-propagated pressure field at point $\mathbf{K}$. This transducer is fixed on a stepping motor in order to build the spatial map of the TR field around the initial source location. As expected, when $\mathbf{K} = \mathbf{S}$, a short pulse emerges that corresponds to the time-reversed initial pulse (cf.~Fig.~\ref{fig:RT_time_foc} and \ref{fig:RT_time_foc2}) 
This result is obtained for a single realization of disorder. In order to get an estimator for the average TR field, the experiment can be repeated for many other configurations of the scatterer positions. From a practical point of view, averaging is achieved by translating the disordered medium. After averaging, the sidelobe level is decreased but the shape of the compressed pulse seems unchanged. This result illustrates the self-averaging property of broadband time reversal in random media.

\begin{figure}
        \begin{center}
                \includegraphics{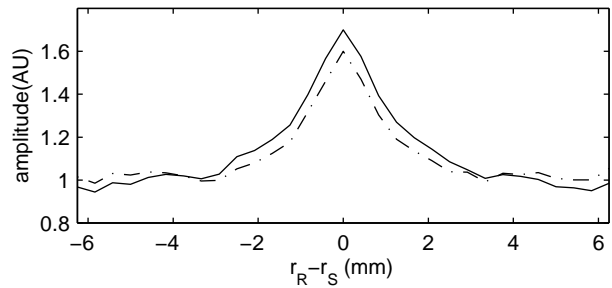}
                \caption{\label{fig:RT_cone2}Full line: averaged time reversal focusing amplitude (at $t=0$) and $\mathbf{K}=\mathbf{S}$ versus distance between the initial source and the TR device. Dotted line: average backscattered intensity versus distance between the source and the TR device.}
        \end{center}
\end{figure}

So far the source position and the time reversal channel were assumed to be far away from each other. Do the TR focusing characteristics change when points $\mathbf{S}$ and $\mathbf{R}$ coincide?
 \begin{figure}
        \begin{center}
        \includegraphics{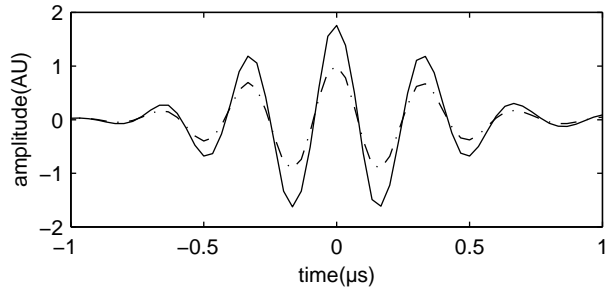}
        \caption{\label{fig:RT_time_foc_enhancement} Average time compression on the initial source ($\mathbf{K}=\mathbf{S}$). Dotted line: the TR device location ($\mathbf{R}$) is far from the initial source point ($\mathbf{S}$), its amplitude is normalized to 1; continuous line: they are at same position ($\mathbf{R}=\mathbf{S}$).}
        \end{center}
\end{figure}
An experimental answer comes from Fig.~\ref{fig:RT_time_foc_enhancement}. When $\mathbf{S}=\mathbf{K}=\mathbf{R}$, the average TR field is almost twice as large compared to the set-up in which $\mathbf{R}$ and $\mathbf{K}=\mathbf{S}$ are far away from each other. Moreover, if the average TR focusing amplitude is measured as a function of distance between points $\mathbf{S}=\mathbf{K}$ and $\mathbf{R}$, the resulting plot looks like the well-known Coherent Backscattering (CBS) peak~(see Fig.~\ref{fig:RT_cone2}).

The CBS is a well-known effect occurring in multiple scattering media, intimately related to the reciprocity property. When a source located at point $\mathbf{S}$ emits a short pulse towards a multiple scattering medium, the average backscattered intensity 
(i.e.,  $\int \left\langle  \Psi(t)^2 \right\rangle dt$ where $\left\langle \bullet \right\rangle$ represents the average over realizations of disorder)
received at the source is twice higher than the one received far away. This is due to the constructive interference between a each path and its reciprocal counterpart, which can only occur at the source. This phenomenon has been observed in many different areas: in optics, in microwaves, in acoustics, etc... It was first predicted by Watson \cite{Watson69}, de~Wolf \cite{deWolf71}, Barabanenkov \cite{Barabanenkov73}. Ten years later, the first experimental evidence of this phenomenon was reported \cite{VanAlbada85,Wolf85}. 

\begin{figure}
\begin{center}
\includegraphics[width=8cm]{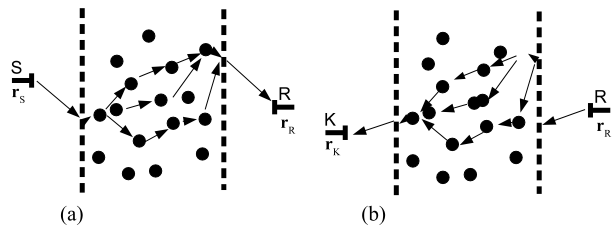}
\caption{\label{fig:schema1}The two steps of a one-channel TR experiment in transmission through a multiple scattering medium. (a) forward propagation step. (b) backward propagation step.}
\end{center}
\end{figure}

\begin{figure}
\begin{center}
\includegraphics[width=7cm]{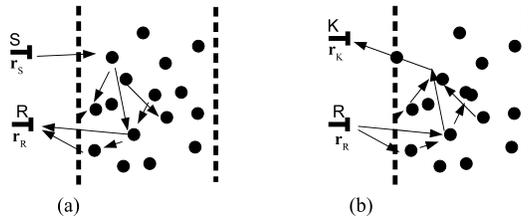}
\caption{\label{fig:schema2}The two steps of a one-channel TR experiment in a backscattering configuration through a multiple scattering medium. (a) forward propagation step. (b) backward propagation step.}        
\end{center}
\end{figure}

Our preliminary results on time reversal focusing (Figs. \ref{fig:RT_time_foc_enhancement} and \ref{fig:RT_cone2}) indicate that there is strong link between TR and CBS. We will later develop a full theoretical analysis to support this statement. Nevertheless, from now on, we propose an intuitive way  to explain the link between CBS and TR based on simple physical arguments. Basically, when the source at $\mathbf{S}$ transmits a Dirac pulse, $\mathbf{R}$ records the Green's function of the medium between point $\mathbf{S}$ and $\mathbf{R}$, $G(\mathbf{S} \rightarrow \mathbf{R}; t)$. This field is time-reversed and $G(\mathbf{S} \rightarrow \mathbf{R},-t)$ is sent back into the medium from point $\mathbf{R}$ to the initial source point, $\mathbf{S}$. The back-propagated field on $\mathbf{S}$ is $G(\mathbf{S} \rightarrow \mathbf{R};-t) \convo G(\mathbf{R} \rightarrow \mathbf{S};t)$. The reciprocity symmetry of the propagation medium implies that the emission and receiving points can be exchanged in the Green's function, i.e.,  $G(\mathbf{S} \rightarrow \mathbf{R};t)=G(\mathbf{R} \rightarrow \mathbf{S};t)$. Therefore the TR field is expressed as $\int G^2(\mathbf{S} \rightarrow \mathbf{R},\tau)d\tau$ at the focusing time ($t=0$). This term is just the integrated intensity of the scattered field received at $\mathbf{R}$ for a source at $\mathbf{S}$. Hence, on average, the dependence of the TR amplitude as a function of the TR device position is nothing else than the backscattered intensity pattern, i.e., the Coherent Backscattering enhancement. These preliminary and qualitative results will be supported rigorously in the rest of the paper. Theoretically we want to determine the average value and the variance of the time reversal field. The theoretical steps are the following:
 \begin{itemize}
\item{Time Reversal in a complex medium on a single realization of disorder.}
\begin{itemize}
        \item Some basics about Green's function formalism in disordered media are reminded
        \item The TR field is deduced from the Green's function formalism.
        \item Its square is calculated
        \end{itemize}
        \item{Averaging over realizations of disorder:}
        \begin{itemize}
        \item General expression for the average TR amplitude in terms of the Vertex function. 
        \item General expression of the variance of the TR amplitude in terms of the Vertex function.
        \end{itemize}
        \item{Approximation of the Vertex: introduction of the Ladder and Crossed diagrams contributions.}
        \begin{itemize}
        \item Analysis of TR in the transmission configuration (cf.~Fig.~\ref{fig:schema1}). 
        \item Analysis of TR in the backscattering configuration (cf.~Fig.~\ref{fig:schema2}).
        \end{itemize}
        \item Discussion about the self-averaging property of TR in a multiple scattering medium.
        \item The link between CBS and TR enhancement is established. 
\end{itemize}
The results will be confirmed by experiments presented in the last part of the paper.
\section{General Theory of Time Reversal in disordered media}
\subsection{Expression of the Green's function inside a multiple scattering medium}
\label{sec:Green}
 In a medium filled with scatterers characterized by a scattering potential $\epsilon$, the Green's function $G(\mathbf{r}_{in} \rightarrow \mathbf{r}_{out};\omega)$ between two points $\mathbf{r}_{in}$ and $\mathbf{r}_{out}$ at frequency $\omega$ is given by\cite{rytov89}:
\begin{eqnarray}
 \label{eq:green_self}
        G(\mathbf{r}_{in} \rightarrow \mathbf{r}_{out};\omega) &=& G_0(\mathbf{r}_{in} \rightarrow \mathbf{r}_{out} ; \omega) \\\nonumber  
        & &+ \int  G_0 (\mathbf{r}_{in} \rightarrow \mathbf{r} ; t) \epsilon(\mathbf{r}) G(\mathbf{r} \rightarrow \mathbf{r}_{out};\omega) d\mathbf{r}
\end{eqnarray}
$G_0$ is the Green's function without scatterers. If this expression is developed, it leads to the well-known Born expansion:
\begin{eqnarray}
 \label{eq:green_decomp}
        G(\mathbf{r}_{in} \rightarrow \mathbf{r}_{out};\omega) = G_0(\mathbf{r}_{in} \rightarrow \mathbf{r}_{out} ; \omega) + \int  G_0 (\mathbf{r}_{in} \rightarrow \mathbf{r} ; t) \epsilon(\mathbf{r}) G_0(\mathbf{r} \rightarrow    \mathbf{r}_{out} ; \omega) d\mathbf{r} \nonumber\\ + \int G_0 (\mathbf{r}_{in} \rightarrow \mathbf{r} ; \omega) \epsilon(\mathbf{r}) G_0 (\mathbf{r} \rightarrow \mathbf{r'} ; \omega) \epsilon(\mathbf{r'}) G_0(\mathbf{r'} \rightarrow \mathbf{r}_{out} ; \omega)  d\mathbf{r}  d\mathbf{r'} + \cdots
\end{eqnarray}
Each term of this development corresponds to an order of scattering: no scattering, single scattering, double scattering, ...
The aim of the multiple scattering theory is to find the average Green's function and its higher moments ($\left\langle G \right\rangle$, $\left\langle G G^{*} \right\rangle$, etc...) from the previous expression. One can prefer to introduce an operator ${\T}$ defined such that\cite{ChengPage68}:
\begin{eqnarray}
        G(\mathbf{r}_{in} \rightarrow \mathbf{r}_{out};\omega) &=& G_{e}(\mathbf{r}_{in} \rightarrow \mathbf{r}_{out} ; \omega) \\\nonumber & & + \int G_{e}(\mathbf{r}_{in} \rightarrow \mathbf{r}_{1} ; \omega) {\T}(\mathbf{r}_{1} \rightarrow \mathbf{r}_{2} ; \omega) G_{e}(\mathbf{r}_{2} \rightarrow     \mathbf{r}_{out} ; \omega) d\mathbf{r}_{2} d\mathbf{r}_{1}
        \label{eq:green_reducible}
\end{eqnarray}
where $G_{e}$ is the average Green's function, i.e., $\left\langle G \right\rangle$. $G_{e}$ is also called the Green's function of the effective medium. This definition implies that $\left\langle {\T} \right\rangle=0$. Nevertheless the next moments of ${\T}$, (e.g. its autocorrelation) are different from zero. Experimentally in acoustics one has access to the time- dependence of the fields rather than their frequency dependence. So we prefer to work directly in the time domain. An inverse Fourier transform of Eq.~(\ref{eq:green_reducible}) yields an expression in the time domain:
\begin{eqnarray}
        G(\mathbf{r}_{in} \rightarrow \mathbf{r}_{out};t) & = & G_{e}(\mathbf{r}_{in} \rightarrow \mathbf{r}_{out} ; t) \\\nonumber
       & &  + \int G_{e}(\mathbf{r}_{in} \rightarrow \mathbf{r}_{1} ; t) \convo {\T}(\mathbf{r}_{1} \rightarrow \mathbf{r}_{2} ; t) \convo G_{e}(\mathbf{r}_{2} \rightarrow   \mathbf{r}_{out} ; t) d\mathbf{r}_{2} \mathbf{r}_{1}
                \label{eq:greenTmaxtrix}
\end{eqnarray}
where $\convo$ is the convolution operator for time variables.
\subsection{Time reversed field}
\label{sec:one_TR}
In this part, we formalize the one-channel TR process in terms of the Green's functions. As seen in the preliminary experiment, the first step of a TR experiment begins by the emission of a short pulse, $f(t)$, around time $t=0$ by the source localized at point $\mathbf{S}$. The recorded field $\Psi$ at point $\mathbf{R}$ is expressed as:
\begin{eqnarray}
        \Psi(t) = G(\mathbf{S} \rightarrow \mathbf{R}; t) \convo f(t)
\end{eqnarray}
From Eq.~(\ref{eq:greenTmaxtrix}), we have:
\begin{eqnarray}
\label{eq:psi}
         \Psi(t) & = & G_{e}(\mathbf{S} \rightarrow \mathbf{R} ; t) \convo f(t)  \nonumber\\
      & & + \left[ \int{G_{e}(\mathbf{S} \rightarrow \mathbf{r}_{1} ; t) \convo {\T}(\mathbf{r}_{1} \rightarrow \mathbf{r}_{2} ; t) \convo G_{e}(\mathbf{r}_{2} \rightarrow     \mathbf{R} ; t) d\mathbf{r}_{1} d\mathbf{r}_{2}} \right] \convo f(t)
\end{eqnarray}
In the following, the first term on the right hand side of Eq. (\ref{eq:psi}) will be neglected. This is fully justified in a backscattering configuration: the effective Green's function has no specular reflection at the interface because the scatterers density is weak\cite{Waterman61} and the ambient fluid is the same in the scattering region and outside. In transmission, this term can also be neglected when the medium is thicker than several elastic mean free paths because the effective Green's function  decays exponentially with depth in the multiple scattering medium. In other words, in thick multiple scattering media, almost all the incoming wave is scattered at least once while traveling across the medium.  Without the first term on the right hand side in Eq.~(\ref{eq:psi}), the propagation from $\mathbf{S}$ to $\mathbf{R}$ can be seen as a three-step process: first the propagation in the effective homogeneous medium from $\mathbf{S}$ to $\mathbf{r}_{1}$, secondly the propagation inside the multiple scattering medium between $\mathbf{r}_{1}$ and $\mathbf{r}_{2}$ and finally the propagation from $\mathbf{r}_{2}$ to $\mathbf{R}$, along all possible paths from $\mathbf{r}_{1}$ to $\mathbf{r}_{2}$ within the scattering region.

Next, we select a part of the scattered signal, referred to as the time-reversal window (TRW). This equals 
\begin{eqnarray}
e(t)=  A \Psi(t) W \left( \frac{t-t_0}{\Delta t} \right)
\label{eq:TRwindow}
\end{eqnarray}
$W$ is the rectangle function \footnote{$W(x)=1$ if $\left|x\right|<1/2$ and  $W(x)=0$ if $\left|x\right|>1/2$}.
$A$ is an amplitude factor that takes into account a possible amplification, but can be set to unity without loss of generality. 
The TRW is centered at time $t_0$ and has a duration $\Delta T$. The selected signal $e(t)$ is time-reversed and transmitted back by point $\mathbf{R}$ through the medium and the field $\Psi_{RT}$ is recorded at point $\mathbf{K}$ (see Fig.~\ref{fig:schema1}b and~\ref{fig:schema2}b):
\begin{eqnarray}
\label{eq:TR_one_realization}
        \Psi_{RT}(t) =   e(-t) \convo G(\mathbf{R} \rightarrow \mathbf{K};t)
\end{eqnarray}
Combining Eq.~(\ref{eq:psi}), Eq.~(\ref{eq:TRwindow}) and Eq.~(\ref{eq:TR_one_realization}), we obtain
\begin{eqnarray}
   \label{eq:TR_field}
         \Psi_{RT}(t) & = &  \int \gamma ( \mathbf{r}_{1},\mathbf{r}_{2},\mathbf{r}_{3},\mathbf{r}_{4} ; t)
         \convo f(-t)  \nonumber\\ & & \convo \underbrace{  {\T}(\mathbf{r}_{3} \rightarrow \mathbf{r}_{4} ; t) \convo \left[ W\left(\frac{-t-t_0}{\Delta T} \right) {\T}(\mathbf{r}_{1} \rightarrow \mathbf{r}_{2} ; -t) \right] }_{\chi(t)}  d\mathbf{r}_{1} d\mathbf{r}_{2} d\mathbf{r}_{3} d\mathbf{r}_{4}  
\end{eqnarray}
where
\begin{eqnarray}
\gamma(\mathbf{r}_{1},\mathbf{r}_{2},\mathbf{r}_{3},\mathbf{r}_{4} ; t) & = & G_{e}(\mathbf{S} \rightarrow \mathbf{r}_{1} ; -t ) \convo G_{e}(\mathbf{R} \rightarrow \mathbf{r}_{3} ; t) \\\nonumber & &  \convo G_{e}(\mathbf{r}_{2} \rightarrow \mathbf{R}; -t) \convo G_{e}(\mathbf{r}_{4} \rightarrow        \mathbf{K}; t)
\end{eqnarray}
To obtain Eq.~(\ref{eq:TR_field}) we have assumed that $\Delta T$ is larger than the duration of $\gamma$. $\gamma$ contains all the effective Green's functions and $\chi$ takes into account propagation inside the disordered medium and the choice of the TRW. Expression (\ref{eq:TR_field}) for the time-reversed field in terms of $G_{e}$ and the ${\T}$ matrix will serve us as a starting point to work out the average and the variance of the TR field.
\subsection{Average TR wavefield}
\label{sec:meanTR}
In Eq.~(\ref{eq:TR_field}), the randomness is contained only in $\chi(t)$. Hence, $\left\langle \chi(t) \right\rangle$ is the key to $\left\langle \Psi_{RT}\right\rangle$. In Appendix~\ref{appendix:calculs} it is shown that:
\begin{eqnarray}
        \left\langle \chi(t) \right\rangle = \left[ \int C_T(\mathbf{r}_{1} , \mathbf{r}_{2} ;\mathbf{r}_{3} , \mathbf{r}_{4};\tau)  W \left(\frac{\tau-t_0}{\Delta T} \right) d\tau \right]  \delta(t)
        \label{eq:chimean}
\end{eqnarray}
where 
\begin{eqnarray}
 C_T(\mathbf{r}_{1} , \mathbf{r}_{2} ;\mathbf{r}_{3} , \mathbf{r}_{4};\tau)= \left\langle {\T}(\mathbf{r}_{1} \rightarrow \mathbf{r}_{2} ; \tau) {\T}(\mathbf{r}_{3} \rightarrow \mathbf{r}_{4} ; \tau)  \right\rangle
\end{eqnarray}
To obtain Eq. (\ref{eq:chimean}) we assumed that in the frequency domain the correlation properties of the ${\T}$-operator only depend on the frequency difference. In the multiple scattering theory $C_T$ is also referred to as the Intensity ``Vertex'' function that links the entry positions $\mathbf{r}_{1}$, $\mathbf{r}_{3}$ to the output ones $\mathbf{r}_{2}$, $\mathbf{r}_{4}$.

We conclude that the average TR field is intimately linked to the Intensity Vertex according to
{
\begin{eqnarray}
        \langle \Psi_{RT}(t) \rangle = 
   \int \gamma(\mathbf{r}_{1},\mathbf{r}_{2},\mathbf{r}_{3},\mathbf{r}_{4} ; t) \convo f(-t) \times
         \left[ \int C_T(\mathbf{r}_{1} , \mathbf{r}_{2} ;\mathbf{r}_{3} , \mathbf{r}_{4};\tau)  W \left( \frac{\tau-t_0}{\Delta T} \right) d\tau \right] \nonumber\\ 
           d\mathbf{r}_{1} d\mathbf{r}_{2} d\mathbf{r}_{3} d\mathbf{r}_{4}  
           \label{eq:meanTRW}
\end{eqnarray}
}
Upon introducing the $\xi$ function
{
\begin{eqnarray}
\xi_{\tau}(\mathbf{X}_1,\mathbf{X}_2,\mathbf{X}_3,\mathbf{X}_4;t) &= & \int  G_{e}( \mathbf{X}_1 \rightarrow  \mathbf{r}_{1}  ; -t )  \convo G_{e}(\mathbf{r}_{2} \rightarrow   \mathbf{X}_2; -t) \nonumber\\ & & \convo  G_{e}(  \mathbf{X}_3 \rightarrow \mathbf{r}_{3} ; t)    \convo G_{e}(\mathbf{r}_{4} \rightarrow      \mathbf{X}_4; t)  \convo f(-t) \nonumber\\& &  \times C_T(\mathbf{r}_{1},\mathbf{r}_{2},\mathbf{r}_{3},\mathbf{r}_{4};\tau) 
          d\mathbf{r}_{1} d\mathbf{r}_{2} d\mathbf{r}_{3} d\mathbf{r}_{4} 
          \label{eq:def_zeta}
\end{eqnarray}
}
Eq.~(\ref{eq:meanTRW}) reads:
\begin{eqnarray}
\label{eq:meanRTzeta}
\langle \Psi_{RT}(t) \rangle = \int \xi_{\tau}(\mathbf{S},\mathbf{R},\mathbf{R},\mathbf{K};t) W(\frac{\tau-t_0}{\Delta T}) d\tau
\end{eqnarray}
In the following, we will see that the $\xi$ function plays a central role in the multiple scattering theory applied to TR.

From a practical point of view, we often deal with a short Time Reversal Window. When its duration $\Delta T$ is sufficiently small so that $C_T(\mathbf{r}_{1} , \mathbf{r}_{2} ;\mathbf{r}_{3} , \mathbf{r}_{4};\tau)$ is almost constant within this interval, the average TR field can be expressed as:
\begin{eqnarray}
         \left\langle \Psi_{RT}(t) \right\rangle \operatorname*{=}_{\Delta T \rightarrow 0}  \Delta T \xi_{t_0}(\mathbf{S},\mathbf{R},\mathbf{R},\mathbf{K},t) 
\end{eqnarray}
The other interesting limit is when the Time Reversal Window contains the whole recorded signal $\Psi(t)$, in which case : 
\begin{eqnarray}
         \left\langle \Psi_{RT}(t) \right\rangle \operatorname*{=}_{\Delta T \rightarrow \infty}  \int \xi_{\tau}(\mathbf{S},\mathbf{R},\mathbf{R},\mathbf{K},t) d\tau      
\end{eqnarray}

So far the average TR field has been worked out. To this end, we have assumed that the first term of the right hand side of Eq.~(\ref{eq:psi}) is negligible. If we do not neglect this term, the term $G_{e}(\mathbf{R} \rightarrow \mathbf{K} ; t)  \convo G_{e}(\mathbf{S} \rightarrow \mathbf{R} ; -t) \convo f(-t)$ must be added to Eq.~(\ref{eq:TR_field}) as well as in the subsequent equations.
\label{sec:meansquared}
To determine the variance of the TR field, we have to average the squared time-reversed field. The coherent terms are neglected under the same conditions as those explained in the previous section. Therefore the average squared field is:
\begin{eqnarray}
        \langle \Psi_{RT}(t)^2 \rangle =   \int \left[ \gamma(\mathbf{r}_{1},\mathbf{r}_{2},\mathbf{r}_{3},\mathbf{r}_{4} ; t-\tau) \convo f(\tau-t) \textstyle \right]  \left[ \gamma(\mathbf{r}_{1}',\mathbf{r}_{2}',\mathbf{r}_{3}',\mathbf{r}_{4}' ; t-\tau') \convo f(\tau'-t) \right]
    \nonumber\\      
\times \left\langle \underbrace{ 
\begin{array}{l}        
{\left( {\T}(\mathbf{r}_{3} \rightarrow \mathbf{r}_{4} ; \tau) \convo  \left[ W(\frac{-\tau-t_0}{\Delta T} ) {\T}(\mathbf{r}_{1} \rightarrow \mathbf{r}_{2} ; -\tau) \right] \right)}
\cr \times                       
{\left( {\T}(\mathbf{r}_{3'} \rightarrow \mathbf{r}_{4'} ; \tau') \convo \left[ W(\frac{-\tau'-t_0}{\Delta T} ) {\T}\left(\mathbf{r}_{1'} \rightarrow \mathbf{r}_{2'} ; -\tau' \right) \right] \right)}   
\end{array}
   }_{\textstyle \chi_2(\tau,\tau')}  \right\rangle\nonumber\\
          \times d\mathbf{r}_{1} d\mathbf{r}_{2} d\mathbf{r}_{3} d\mathbf{r}_{4} d\mathbf{r}_{1'} d\mathbf{r}_{2'} d\mathbf{r}_{3'} d\mathbf{r}_{4'} d\tau d\tau'
          \label{eq:def_variance}
\end{eqnarray}
All effects induced by multiple scattering are governed by the function $\chi_2$. Obviously the  $\chi_2$ directly depends on the eight-point function in the frequency domain, $\left\langle {\T}(\mathbf{r}_{3} \rightarrow \mathbf{r}_{4} ; \omega) {\T}^{*}(\mathbf{r}_{1} \rightarrow \mathbf{r}_{2} ; \omega) {\T}(\mathbf{r}_{3'} \rightarrow \mathbf{r}_{4'} ; \omega') {\T}^{*}(\mathbf{r}_{1'} \rightarrow \mathbf{r}_{2'} ; \omega')  \right\rangle$. In the following, we assume that the Thouless number of our medium is much larger than one so that all the random fields are jointly Gaussian and the cumulant theorem can be applied. Thus the eight-field correlation function can be expressed in terms of the four-field correlations. This approximation means that the probability for having crossing paths in the medium is weak\cite{Berkovits94}.  The full calculation presented in Appendix~\ref{appendix:calculs} is tedious but without particular difficulty. Finally it comes:
\begin{eqnarray}
        \langle \Psi_{RT}(t)^2 \rangle - \langle \Psi_{RT}(t) \rangle^2  =  \int \left[ \xi_{-t'} (\mathbf{S},\mathbf{R},\mathbf{S},\mathbf{R},\tau) W^2\left(\frac{-t'-t_0}{\Delta T}\right) \right] \convo_{t'=t} \xi_{t'}(\mathbf{K},\mathbf{R},\mathbf{K},\mathbf{R},\tau) d\tau \nonumber\\
        +  \left[ \xi_{t'} (\mathbf{S},\mathbf{R},\mathbf{R},\mathbf{K},\tau) W\left (\frac{t'-t_0}{\Delta T} \right) \right]  \convo_{t'=t} \convo_{\tau=2 t} \left[\xi_{-t'} (\mathbf{S},\mathbf{R},\mathbf{R},\mathbf{K},\tau) W\left( \frac{-t'-t_0}{\Delta T} \right) \right]
        \label{eq:varRTzeta}
\end{eqnarray}
 The  subscripted $\convo$ operator notation is introduced in order to remove any possible ambiguity.  $f(t') \convo_{t'=t} g(t') = \int_{-\infty}^{\infty} f(t') g(t-t') dt'$. In the short and long TR window limits respectively, Eq.~(\ref{eq:varRTzeta}) simplifies to:
\begin{eqnarray}
        \langle \Psi_{RT}(t)^2 \rangle - \langle \Psi_{RT}(t) \rangle^2   \operatorname*{=}_{\Delta T \rightarrow 0}   \Delta T \int \xi_{t_0} (\mathbf{S},\mathbf{R},\mathbf{S},\mathbf{R},\tau) \xi_{t_0+t}(\mathbf{K},\mathbf{R},\mathbf{K},\mathbf{R},\tau) d\tau \nonumber\\
        + \Delta T \xi_{t_0} (\mathbf{S},\mathbf{R},\mathbf{R},\mathbf{K},\tau) \convo_{\tau=2 t} \xi_{t_0} (\mathbf{S},\mathbf{R},\mathbf{R},\mathbf{K},\tau)
        \label{eq:varRTzetaShortApprox}
\end{eqnarray}
        \begin{eqnarray}
        \langle \Psi_{RT}(t)^2 \rangle  -  \langle \Psi_{RT}(t) \rangle^2 \rangle \operatorname*{=}_{\Delta T \rightarrow \infty}    \int \int  \xi_{-t'} (\mathbf{S},\mathbf{R},\mathbf{S},\mathbf{R},\tau)  \xi_{t'}(\mathbf{K},\mathbf{R},\mathbf{K},\mathbf{R},\tau) d\tau dt' \nonumber\\
        + \int \xi_{t'} (\mathbf{S},\mathbf{R},\mathbf{R},\mathbf{K},\tau)  \convo_{\tau=2 t} \xi_{-t'} (\mathbf{S},\mathbf{R},\mathbf{R},\mathbf{K},\tau) dt' 
        \label{eq:varRTzetaLongApprox}
\end{eqnarray}
\subsection{Approximation of the Vertex}
\label{sec:approxVertex}
\begin{figure}
        \begin{center}
                \includegraphics{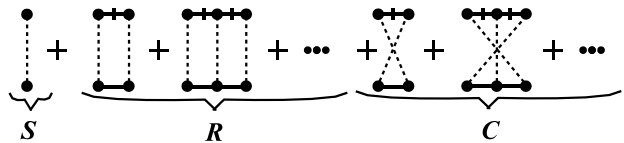}
                \caption{\label{fig:ladders}Diagrammatic representation of the expansion of the Vertex: (${S}$) represents the single scattering contribution, (${R}$) the ``Ladder diagrams'' and ($C$) the most-crossed diagrams that restore reciprocity}
        \end{center}
\end{figure}

The Vertex can be split into 2 distinct contributions\cite{Tiggelen99}: the irreducible one ($U$) and the reducible one (${R}$). 
\begin{eqnarray}
\langle {\T}(\mathbf{r}_{1} \rightarrow \mathbf{r}_{2} ; t) {\T}(\mathbf{r}_{1'} \rightarrow \mathbf{r}_{2'} ;t) \rangle= U(\mathbf{r}_{1} , \mathbf{r}_{2},\mathbf{r}_{1'} , \mathbf{r}_{2'};t)  + R(\mathbf{r}_{1} , \mathbf{r}_{2},\mathbf{r}_{1'} , \mathbf{r}_{2'};t) 
\label{eq:ExpensionVertex}
\end{eqnarray}
The ``Boltzmann approximation'' consists in replacing ($U$) by the scattering by one scatterer (here noted ${S}$). This low density approximation turns (${R}$) into an incoherent multiple scattering series called 'the Ladder diagrams'' which obeys the radiative transfer equation. (${R}$) itself does not obey reciprocity. Therefore the Boltzmann approximation fails in describing any reciprocity-dependent effect, such as Coherent Backscattering. Beyond the Boltzmann approximation,  reciprocity is restored by adding ``most-crossed diagrams''($C$), which are irreducible, to the diagrammatic expansion (see Fig.~\ref{fig:ladders}).
Obviously, due to the straightforward link between the Vertex function and the $\xi$ function, (see Eq.~(\ref{eq:def_zeta})), the same decomposition appears for $\xi$:
\begin{eqnarray}
        \xi_{\tau} (\mathbf{X}_1,\mathbf{X}_2,\mathbf{X}_3,\mathbf{X}_4,t) &=& \xi^S_{\tau} (\mathbf{X}_1,\mathbf{X}_2,\mathbf{X}_3,\mathbf{X}_4,t) \nonumber\\ & &+ \xi^R_{\tau} (\mathbf{X}_1,\mathbf{X}_2,\mathbf{X}_3,\mathbf{X}_4,t) + \xi^{C}_{\tau} (\mathbf{X}_1,\mathbf{X}_2,\mathbf{X}_3,\mathbf{X}_4,t)
        \label{eq:zeta_decomp}
\end{eqnarray}
In the following, we shall neglect the single scattering contribution. The reciprocity symmetry means that all Green's functions are invariant upon permuting source and receiver positions~: ($G(\mathbf{r}_{1} \rightarrow \mathbf{r}_{2} ; t)$ = $G(\mathbf{r}_{2} \rightarrow \mathbf{r}_{1} ; t)$). Consequently, in reciprocal media, a strong relation exists between the Ladder and the most-crossed diagrams; $R(\mathbf{r}_{1},\mathbf{r}_{2} ,\mathbf{r}_{2'} , \mathbf{r}_{1'};t)=C(\mathbf{r}_{1} , \mathbf{r}_{2},\mathbf{r}_{1'} , \mathbf{r}_{2'};t)$. The same relation holds for $\xi^{C}$ and $\xi^{R}$
\begin{eqnarray}
\xi^C_{\tau}(\mathbf{X}_1,\mathbf{X}_2,\mathbf{X}_3,\mathbf{X}_4,t)=\xi^R_{\tau}(\mathbf{X}_1,\mathbf{X}_2,\mathbf{X}_4,\mathbf{X}_3,t) 
\label{eq:zeta_reciprocity}
\end{eqnarray}
The solution for the total Vertex $C_T(\mathbf{r}_{1} , \mathbf{r}_{2} ;\mathbf{r}_{3} , \mathbf{r}_{4};\tau)$ can be written $F(\mathbf{r}_{1} , \mathbf{r}_{2} ; t)  \delta(\mathbf{r}_{1} - \mathbf{r}_{3}) \delta(\mathbf{r}_{2}-\mathbf{r}_{4})$. Under the diffusion approximation,  $F(\mathbf{r}_{1} , \mathbf{r}_{2} ; t)$ obeys the diffusion equation\cite{Mark88}. Using Eq.~(\ref{eq:def_zeta}) gives
\begin{eqnarray}        
\xi^R_{\tau}(\mathbf{X}_1,\mathbf{X}_2,\mathbf{X}_3,\mathbf{X}_4,t) =   \int G_{e}(\mathbf{X}_1 \rightarrow \mathbf{r}_{1} ; -t )  \convo G_{e}(\mathbf{r}_{2} \rightarrow      \mathbf{X}_2; -t) \nonumber\\   \convo G_{e}(\mathbf{X}_3 \rightarrow \mathbf{r}_{1} ; t) \convo  G_{e}(\mathbf{r}_{2} \rightarrow \mathbf{X}_4; t)  \convo f(-t)  F(\mathbf{r}_{1},\mathbf{r}_{2} ; t_0 ) d\mathbf{r}_{1} d\mathbf{r}_{2}
\label{eq:zeta_Ladder_Approx}  
\end{eqnarray}.
This expression for $\xi$ is valid in the diffusion approximation and takes into account the reciprocity. The next two sections of the article will be devoted to elucidating the respective role of the reducible and the most-crossed components of the vertex function in TR focusing. First we will consider the transmission configuration and next the backscattering configuration which is more complicated.
\section{Transmission configuration}
\subsubsection{Average TR field}
\label{sec:ave_TR_field}
\begin{figure}[htbp]
        \begin{center}
  \includegraphics{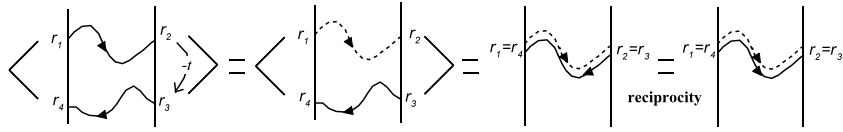}
        \caption{\label{fig:meanfieldcontribution}Paths that contribute to the TR mean field in transmission configuration. Dotted line indicates conjugated path.}        
\end{center}
\end{figure}

In transmission, the points $\mathbf{S}$ and $\mathbf{R}$ are on opposite sides of the multiple scattering medium. The consequence is that $\xi^R_{\tau}(\mathbf{S},\mathbf{R},\mathbf{R},\mathbf{K},t)$ is null. Indeed, the effective Green's functions are appreciable only on a skin layer whose thickness is one or a few mean free paths. If $\mathbf{r}_{1}$ is in the skin layer at the sample input, then the effective Green function $G_{e}(\mathbf{R} \rightarrow \mathbf{r}_{1}   ; t)$ is non zero but in this case $G_{e}(\mathbf{S} \rightarrow \mathbf{r}_{1}   ; -t)$ is negligible. Reciprocally, if $G_{e}(\mathbf{R} \rightarrow \mathbf{r}_{1}   ; t)$ is non zero, $G_{e}(\mathbf{S} \rightarrow \mathbf{r}_{1}   ; -t)$ is zero.  Therefore the average TR field (Eq.~(\ref{eq:meanRTzeta})) becomes:
\begin{eqnarray}
\langle \Psi_{RT}(t) \rangle = \int \xi^C_{\tau}(\mathbf{S},\mathbf{R},\mathbf{R},\mathbf{K};t) W(\frac{\tau-t_0}{\Delta T}) d\tau
\label{eq:averageTRfield}
\end{eqnarray}
That is to say that in transmission TR focusing properties come only from the existence of the most-crossed diagrams. A visual interpretation of the effect of averaging over realizations on TR propagation is proposed in Fig.~\ref{fig:meanfieldcontribution}.
The same kinds of conventions as used by Akkermans and Montambaux in ref \cite{Akkermans03} are adopted here. For clarity only propagation inside the multiple scattering medium is represented. An acoustic path starts from $\mathbf{r}_{1}$ and goes out at point $\mathbf{r}_{2}$ with an associated. The wave is time-reversed. The new path starts from $\mathbf{r}_{3}$ and exits at $\mathbf{r}_{4}$ with another phase.  Time parity transformation is formally equivalent to replace the forward path by its conjugate (see Fig. \ref{fig:meanfieldcontribution}). The average process cancels out most of these paths. The only remaining paths are those that propagate along the same trajectory during the forward and the backward steps. (i.e. $\mathbf{r}_{1} = \mathbf{r}_{4}$ and $\mathbf{r}_{2}=\mathbf{r}_{3}$, see Fig. \ref{fig:meanfieldcontribution}). . Moreover the path and its conjugate have opposite propagation direction. It means that these paths will constructively interfere only if one of them can be replaced by its reciprocal. Thus the focusing only originates from the most-crossed diagrams that result from the reciprocity property. As the pairs of paths are linked by a convolution operator, they interfere constructively only at time $t=0$. Usually in this kind of representation, a simple product operator replaces the convolution operator since one works in the frequency domain. Then constructive interferences occur at any time.\\
Therefore in transmission and on average, the TR focusing only succeeds thanks to reciprocity. In other words, only the most crossed-diagrams contribute to the mean TR field. Usually the most-crossed diagrams are thought to be significant only in backscattering; here we have shown that they play a fundamental role in transmission as well when a time reversal experiment is performed. Now let us consider the statistical fluctuations around the mean value. 
\subsubsection{Variance of the TR field}
\begin{figure*}
\begin{center}
\includegraphics{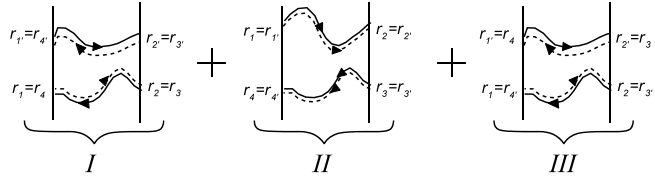}
\caption{\label{fig:variancecontribution}Paths that contribute to the average squared TR field.}   
\end{center}
\end{figure*}

Using the reciprocity principle (Eq.~(\ref{eq:zeta_reciprocity})) and the observation that in transmission $\xi^R_{\tau}(\mathbf{S},\mathbf{R},\mathbf{R},\mathbf{K},t)$ vanishes, one infers that:
\begin{eqnarray}
\xi^C_{\tau}(\mathbf{S},\mathbf{R},\mathbf{K},\mathbf{R},t)= 0 
\label{eq:zeta_transmission2}
\end{eqnarray}
Thus the second term of the variance in Eq. (\ref{eq:varRTzeta}) only originates from the Ladder contribution while the third term comes from the most-crossed diagrams. Again, a graphical interpretation is proposed in Fig.~\ref{fig:variancecontribution}.
In the symbolic equation represented in Fig.~\ref{fig:variancecontribution}, there are now four paths in each term: two correspond (continuous line) to the forth and back propagation in the TR process and the other two symbolize the conjugate paths required to compute the square of the TR field. The three terms on the right hand side of the figure correspond to the 3 terms on the right hand side of Eq.~(\ref{eq:varRTzeta}). The first one denotes the square of the average value of the TR field. The second one corresponds to the contribution of the forward path (from $\mathbf{r}_{1}$ to $\mathbf{r}_{2}$) with its conjugate together with the backward (from $\mathbf{r}_{3}$ to $\mathbf{r}_{4}$) path and its conjugate. The propagation directions of paths in each pair are the same. This means that the second contribution does not involve reciprocity of the medium. These two paths are linked together by a simple product operator, and is therefore a background contribution that slowly changes with time. The third term due to reciprocity only appears at the focusing ($t\approx0$).   
\section{Backscattering configuration}
In the backscattering configuration, $\xi^R_{\tau}(\mathbf{S},\mathbf{R},\mathbf{R},\mathbf{K},t)$ can be non zero when $\mathbf{S}$ and $\mathbf{K}$ are both close to $\mathbf{R}$. In order to illustrate the theoretical results obtained in the backscattering configuration and  to compare them later to experiment, we will consider an explicit solution for the $\xi$ function. In the general case, the computation of $\xi$ is very complex. However, it can be greatly simplified if we assume that all active elements ($\mathbf{S}$, $\mathbf{R}$ and $\mathbf{K}$) are in the far field zone of the multiple scattering medium and that the initial pulse, $f(t)$ has a bandwidth much smaller than its central frequency. Under these conditions, $\xi$ is written as:
\begin{eqnarray}
\xi_{\tau}(\mathbf{X}_1,\mathbf{X}_2,\mathbf{X}_3,\mathbf{X}_4;t) & = &[ {\tilde{F'}(\mathbf{X}_4 - \mathbf{X}_1,\mathbf{X}_3 - \mathbf{X}_2; \tau)} \nonumber \\ & & +  {\tilde{F'}(\mathbf{X}_3 - \mathbf{X}_1,\mathbf{X}_4 - \mathbf{X}_2; \tau)}  ] f(-t)
\label{eq:zetaFT2}
\end{eqnarray}
where $\tilde{F'}$ is the 2D-spatial Fourier transform of $F'$ which is a function directly linked to $F$ (Appendix \ref{sec:far_field_approx}) introduced in Eq.~(\ref{eq:zeta_Ladder_Approx}) as a solution of the diffusion equation with boundary conditions and calculated recently\cite{Tourin99a}. The computation is quite long and complex and is out of the scope of this article. We have applied these results using parameters that correspond to our experimental ultrasonic experiment performed in a water tank. The central frequency of the initial pulse is $3.5\:MHz$, the sound speed in the homogeneous medium is $c=1.5\:mm/\mu s$, the thickness of the multiple scattering medium is $35\:mm$. The diffusion coefficient is $D=3.2\:mm^2/s$, the elastic mean free path is $\ell=4\:mm$ and the transport mean free path is $\ell^{*}=4.8\:mm$\cite{Tourin00}.   
\subsubsection{Average TR field}
\label{sec:diff_mean}
\begin{figure}[hbtp]
\begin{center}
                \includegraphics[]{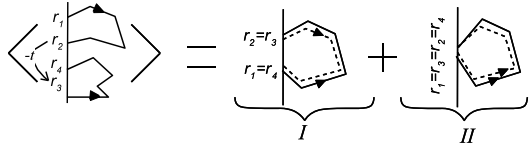}
                \caption{\label{fig:meanfieldCone}Paths contributing to the average TR field in backscattering}
\end{center}
\end{figure}

\begin{figure}
\begin{center}
\includegraphics[width=8cm]{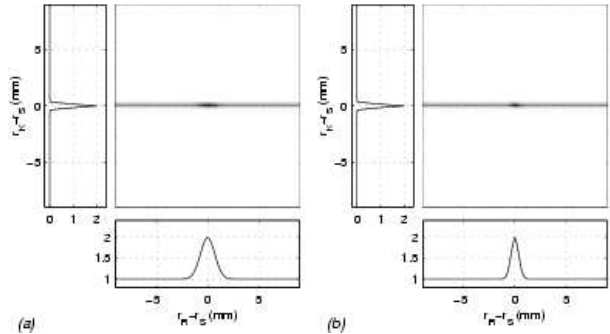}
\end{center}
\caption{\label{fig:simul_ave_spatio_spatio}
Numerical computation of the spatial dependence of the average TR wavefield at time $t=0$ versus the distance between the source and the measurement points (vertical scale) and the distance between TR channel and the initial source (horizontal scale). The representations are in linear gray scale. The vertical curve is the plot of the field for $\mathbf{R}=\mathbf{S}$  and the horizontal one for $\mathbf{K}=\mathbf{S}$. (a) and (b) correspond to TR focusing for time reversal windows respectively centered at $15\:\mu s$ and $55\:\mu s$. }
\end{figure}

\begin{figure}
        \begin{center}
        \includegraphics{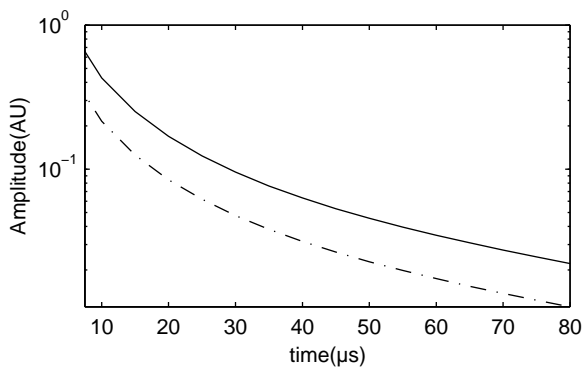}
        \end{center}
        \caption{\label{fig:simul_tof} Evolutions of the average time-reversed peak amplitude recorded at the source ($\mathbf{K}=\mathbf{S}$) with respect to the center of the TR window (time $t_0$). Continuous line: TR channel is at the same position as the source ($\mathbf{S}=\mathbf{R}=\mathbf{K}$). Dashed line: $\mathbf{S}$ and $\mathbf{R}$ are far away from each other.
}
\end{figure}

From Eq.~(\ref{eq:meanRTzeta}) and Eq.~(\ref{eq:zeta_decomp}) show that the average time reversed field involves now both the ladder and the most-crossed diagram contributions. They are represented in terms of paths in Fig.~\ref{fig:meanfieldCone}.
The contribution \textit{I} due to the crossed diagrams corresponds to the one previously considered in transmission. However when the four points $\mathbf{r}_i$ are at the same location, the second contribution (\textit{II}) adds up to the previous one. It describes constructive interferences between a path and its conjugate that follows the same sequence of scattering events in the same order. As the direction of propagation is the same along both paths, this contribution would survive even if the reciprocity was broken. Both contributions show up only around time $t=0$. The spatial focusing including these two terms is plotted in Fig.~\ref{fig:simul_ave_spatio_spatio}.
The ladder diagrams (contribution II in Fig.~\ref{fig:meanfieldCone}) only contribute when $\mathbf{S}$, $\mathbf{R}$ and $\mathbf{K}$ are sufficiently close from each other.  The reciprocity principle links each ladder diagram to a crossed one. Therefore the amplitude of the TR peak is twice higher when the TR channel and the source coincide (see Fig.~\ref{fig:simul_tof}). This behavior is very similar to the Coherent Backscattering Effect(CBS). More details are given in section~\ref{sec:cone}. A surprising conclusion can be already drawn : when the reciprocity property of the propagation medium is broken, for instance by a strong flow, TR focusing still works as long as the initial source and the TR channel are at the same position. This is contrary to the intuitive idea that time reversal requires at least reciprocity of the medium.
\subsubsection{Variance of the TR field}
The expression of the variance of $\Psi_{RT}$ is quite complex. Indeed, the variance is composed of not less than 8 terms. Nevertheless in the far field limit and for short TR windows, its expression is greatly simplified (Eq.~(\ref{eq:varRTzetaShortApprox}) and Eq.~(\ref{eq:zetaFT2})):
{
\begin{eqnarray}
\label{eq:var_far_field_ladder_cross}
        \langle \Psi_{RT}(t)^2 \rangle - \langle \Psi_{RT}(t) \rangle^2 \operatorname*{=}_{\Delta T \rightarrow 0}   ( \tilde{F'}(0,0; t_0) + \tilde{F'}(\mathbf{R}-\mathbf{S},\mathbf{R}-\mathbf{S}; t_0) )  \nonumber\\
         \times  ( \tilde{F'}(0,0; t_0+t) + \tilde{F'}(\mathbf{R}-\mathbf{K},\mathbf{R}-\mathbf{K}; t_0+t) ) 
  \Delta T \int f(\tau)^2 d\tau  \nonumber\\
        +  \left[ \tilde{F'}(\mathbf{K}-\mathbf{S},0; t_0)+\tilde{F'}(\mathbf{R}-\mathbf{S},\mathbf{K}-\mathbf{R}; t_0) \right]^2 \Delta T \left[ f(-t) \convo f(-t) \right](2 \: t)
\end{eqnarray}
}
The variance reduces to the sum of two terms. The first one represents a slowly varying contribution. If the source and the TR channel positions are identical ($\mathbf{R}=\mathbf{S}$) this term is twice higher than if they would have been far away from each other. The same conclusion applies when the TR channel and the recording point are identical ($\mathbf{R}=\mathbf{K}$). Thus when all three points coincide, we expect the variance to be enhanced by a factor four.

\begin{figure*}
        \begin{center}
        \includegraphics{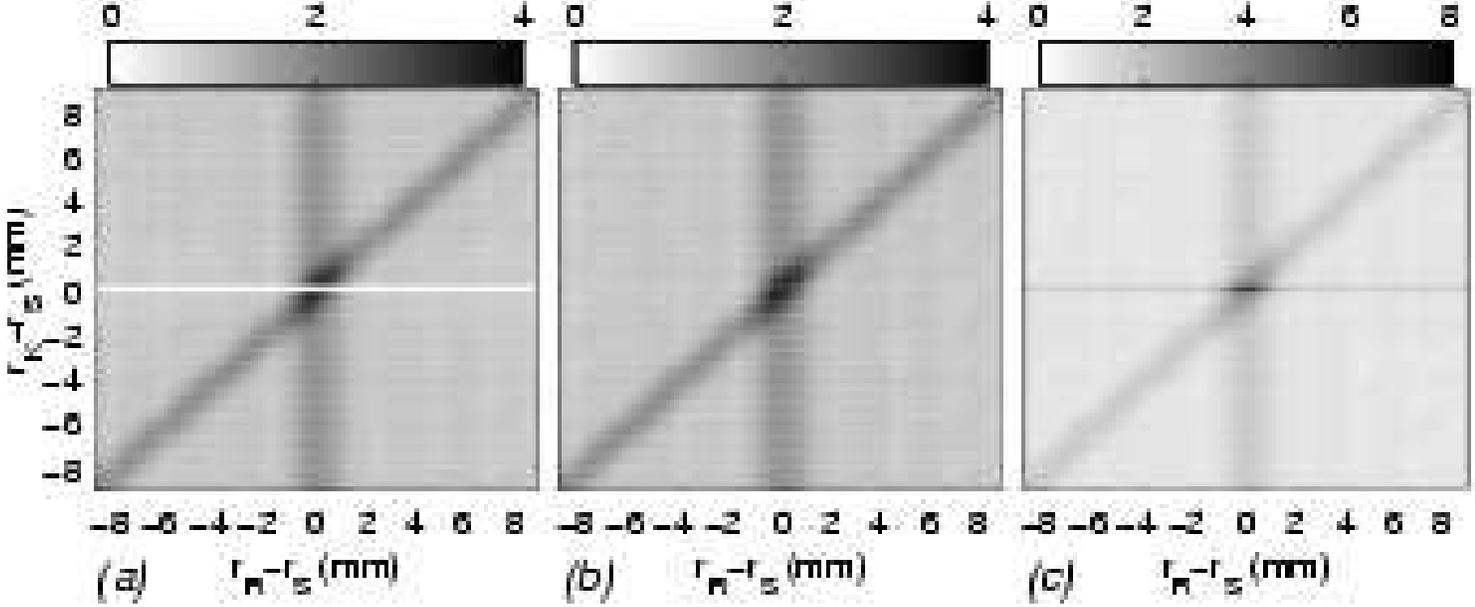}
        \caption{\label{fig:simul_var3} 3 Grey level representations of the variance. The vertical scale is the distance between the observer $\mathbf{K}$ and the source $\mathbf{S}$. The horizontal scale is the distance between the TR channel $\mathbf{R}$ and the initial source $\mathbf{S}$. The time reversal window is centered at $t_0=15\:\mu s$. Figures (a), (b) and (c) represent the variance  at 3 different times for which the oscillating term is respectively minimum, zero and maximum.}
        \end{center}
\end{figure*}

The second term oscillates twice faster than the central frequency of the initial pulse. This term is significant only at the focus location ($\mathbf{S}$) and at the focus time ($t=0$). Moreover when the three points are all at the same location, this term is also enhanced by a factor four. Altogether the variance can be enhanced by a factor up to 8. In Fig.~\ref{fig:simul_var3} is plotted the spatial dependence of the variance at 3 different times for which the oscillating term is respectively minimum, zero and maximum.

In conclusion, we have found three contributions to the variance of the TR field at backscattering. The first one is a spatially uniform background due to the ladder diagrams. In addition to this the most-crossed diagrams are responsible for a ``diagonal enhancement'' ($\mathbf{R}=\mathbf{K}$) and a ``vertical enhancement'' ($\mathbf{R}=\mathbf{S}$) (see Fig.~\ref{fig:simul_var3}). Now we have complete expressions and diagrammatic interpretations for the mean value and the variance of the TR field in transmission as well as in backscattering. In the next two sections, we apply these results to demonstrate  ``hyper-resolution' and ``self-averaging'' within the framework of the self-consistent diagrammatic approach. Finally we establish the link between TR focusing and Coherent Backscattering (CBS).

\section{Hyper-focusing}
One of the most striking effects in one-channel TR inside scattering or reverberant media is the hyper-focusing effect\cite{Dowling92,Draeger97,Derode00,Blomgren02,Tourin99b}. Resolutions of about one wavelength have been experimentally observed. In transmission, we can interpret this effect using  Eqs.~(\ref{eq:zeta_reciprocity}), (\ref{eq:zeta_Ladder_Approx}) and ~(\ref{eq:averageTRfield}). Indeed the average TR field can be seen as a converging field emerging from an adaptive lens with an aperture function given by $O(\mathbf{r}_{1})= \int ( \int F(\mathbf{r}_{1} , \mathbf{r}_{2} ; \tau) W[(\tau-t_0)/\Delta T] d\tau ) \times  G_{e}(\mathbf{R} \rightarrow \mathbf{r}_{2} ; t) \convo G_{e}(\mathbf{r}_{2} \rightarrow     \mathbf{R}; -t) \convo f(-t) d\mathbf{r}_{2}$ where $\mathbf{r}_{1}$ is the coordinate on the virtual lens.  The maximum width of $O(\mathbf{r}_{1})$ is the transverse dimension of the sample. For an infinite slab, half-wavelength resolution is achieved with only a single channel TR! From a practical point of view, if the source is not really point-like, its angular spectrum sets the lower limit for the spatial resolution. In backscattering, the same formalism can be applied.
\section{Self-averaging property of TR}
The time reversed signal at the focusing point ($\mathbf{S}$) consists of a short pulse surrounded by side lobes that fluctuate from one realization of disorder to the other. We define the Signal-to-Noise ratio (SNR) at the focusing location ($\mathbf{K}$=$\mathbf{S}$) as the intensity of the peak (at $t=0$) divided by the variance next to the peak. In other words,  $SNR= \langle \Psi_{RT}(0) \rangle^2 / \langle \Psi_{RT}(t)^2 \rangle$. Giving a general expression of the $SNR$ is quite complex. However if the initial excitation function $f(t)$ is narrow-band and frequency-dependent dissipation is negligible then $G_{e}(\mathbf{S} \rightarrow \mathbf{r}_{1} ; -t )  \convo G_{e}(\mathbf{r}_{1} \rightarrow    \mathbf{S}; t)$~ can be approximated by a Dirac $\delta(t)$. In that case the SNR becomes for short and long time-reversal windows respectively: 
\begin{eqnarray}
        SNR \operatorname*{=}_{\Delta T \rightarrow 0}  \frac{\Delta T f^2(0)}{ \int f^2(t) dt} 
        \label{eq:SNR1}
\end{eqnarray}
and 
\begin{eqnarray}
        SNR \operatorname*{=}_{\Delta T \rightarrow \infty}  \frac{f^2(0)}{ \int f^2(t) dt }  \frac{\left(\int F(\mathbf{r}_{1},\mathbf{r}_{2},\tau) d\mathbf{r}_{1} d\mathbf{r}_{2} d\tau \right) ^2}{\int  F(\mathbf{r}_{1},\mathbf{r}_{2},\tau)^2 d\tau d\mathbf{r}_{1} d\mathbf{r}_{2}}
        \label{eq:SNR2}
\end{eqnarray}
Equation~(\ref{eq:SNR1}) is equal to the ratio  between the duration of the time reversal window and the one of the initial pulse $f(t)$. As for Eq.~(\ref{eq:SNR2}), it represents the ratio of the time spreading of the initial pulse due to diffusion and the pulse duration. Hence for small TR windows the SNR ratio increases linearly with respect to the TR window length and reaches a plateau for large TR windows. This result, established here within multiple scattering theory, has been also found from phenomenological assumptions in several previous papers (e.g. \cite{Derode99}). Expressions (\ref{eq:SNR1}) and (\ref{eq:SNR2}) also show that the SNR increases as the pulse length decreases, i.e. when the bandwidth is enlarged. Broad band time reversal focusing in a complex medium tends to be a self-averaging process: the TR field obtained on one realization of disorder is a good estimator of its average field, provided that frequency-dependent dissipation is neglected. An experimental illustration has been provided in section~\ref{sec:preliminary} (Fig.~\ref{fig:RT_time_foc} and Fig.~\ref{fig:RT_time_foc2}).
\section{Time Reversal at backscattering vs. CBS}
\label{sec:cone}
To record the CBS in an acoustic experiment (\cite{Tourin97,Tourin99a}), a source located at $\mathbf{S}$ sends out a short pulse into a multiple scattering medium. The average reflected intensity is recorded at time $t_0$ for each point $\mathbf{R}$.
\begin{eqnarray}
        I&=&\left\langle G(\mathbf{S} \rightarrow \mathbf{R}; t_0)^2  \right\rangle \nonumber\\
        &\approx& \xi_{t_0}(\mathbf{S},\mathbf{R},\mathbf{S},\mathbf{R},t=0)
        \label{eq:eqcone}
\end{eqnarray}
This expression defines the spatial distribution of the baskscattered intensity at a given time. Interestingly, the average TR field for $\mathbf{S}=\mathbf{K}$ and at $t=0$ reads
\begin{eqnarray}
        \label{eq:meancone}
         \left\langle \Psi_{RT}(t) \right\rangle=  \Delta T \xi_{t_0}(\mathbf{S},\mathbf{R},\mathbf{R},\mathbf{S},t=0) 
\end{eqnarray}
The expression of the average TR field is very close to the expression of the average backscattered intensity (Eq.~\ref{eq:eqcone}). Yet, the exact equality is obtained only if the third and fourth entries are permuted. The physical signification of this permutation is fundamental. It implies that from TR to the CBS the role of the ladder and most-crossed diagrams are exchanged: in the CBS, the ``background'' and the ``enhancement'' intensity are respectively due to the ladder and the most-crossed diagrams, while in the average TR field, the background originates from the most-crossed diagrams and the enhancement from the ladders!

\begin{figure}
        \begin{center}
         \includegraphics[width=8cm]{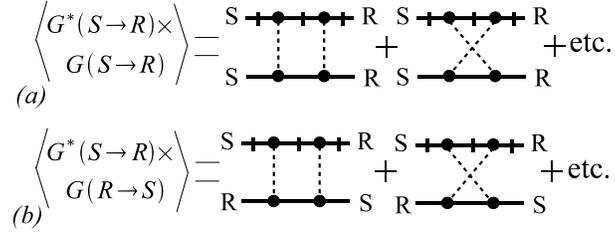}
        \end{center}
        \caption{\label{fig:laddercross}Diagrammatic representation of  the intensity (a) and the time-reversal amplitude at the focal spot (b). On the left part of the diagrams are represented the position of the 2 sources : twice $\mathbf{S}$ in (a), $\mathbf{S}$ and $\mathbf{R}$ in (b). On the right part of the diagrams are represented the positions of the 2 receivers : twice $\mathbf{R}$ in (a), $\mathbf{R}$ and $\mathbf{S}$ in (b). Only second order diagrams are drawn here. The left diagrams corresponds to Ladder contribution and the right ones to most-crossed contribution.}
\end{figure}

This property appears more explicitly in the diagrammatic representations of the average intensity on the one hand and the average TR field amplitude on the other hand  (Fig.~\ref{fig:laddercross}). Let us recall the conventions of diagrammatic representations. First, the sources are on the left part of the diagrams while the receivers are on the right part. Roughly, a dot can be interpreted as a scattering point and  two dots linked by a dashed line represent the same scatterer. A thick segment between two dots or between a dot and the one of the four outer positions symbolizes the effective Green's function; it is barred to indicate complex conjugation. The four outer positions are classified in two pairs : two positions form a pair if they are linked to the same scatterer by two Green's functions, one of them being conjugated. A rule exists to determine whether a diagram is significant or not. A diagram will be significant only if the two positions of each pair are sufficiently close and the contribution will reach its maximum when the two positions are identical.
Hence, in the case of the intensity (Fig.~\ref{fig:laddercross}a), the well-known result is retrieved : the ladders contribute whatever the distance between $\mathbf{R}$ and $\mathbf{S}$ whereas the most-crossed diagrams are only significant when $\mathbf{R} \approx \mathbf{S}$. In the case of TR, the role of diagrams are now exchanged (Fig.~\ref{fig:laddercross}b): the crossed ones always contribute to the focusing amplitude whatever the distance between $\mathbf{R}$ and $\mathbf{S}$. As to the ladders, they only arise when $\mathbf{R} \approx \mathbf{S}$!

A direct consequence is that in a non-reciprocal medium the ``enhancement'' of the TR average field would remain, whereas the background would vanish. On the contrary, the enhancement of the backscattered intensity (i.e., CBS) would disappear, whereas the background would remain.
\section{Experimental results}
The experimental setup was introduced already (see Fig.~\ref{fig:setup}). We now perform a ``dynamic time reversal'', i.e., we do not time reverse the whole signal but only a part of it defined by the Time Reversal Window $W((t-t_0)/\Delta T)$ as in Eq.~(\ref{eq:TRwindow}). The experimental results we present here have been obtained in backscattering, which is the most interesting configuration. First we focus on the average time reversed field, then on its variance. 
\subsection{Average}
\label{exp:ave}

\begin{figure}
        \begin{center}
         \includegraphics[width=8cm]{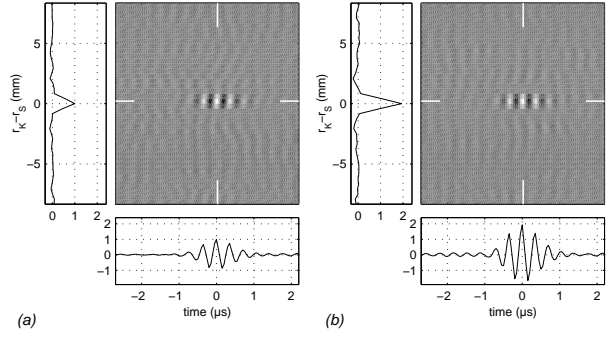}
        \end{center}
        \caption{\label{fig:mean_spatio_tempo_1et2}Spatio-temporal representation of the TR field in gray scale. The vertical axis corresponds to the distance between the measurement point ($\mathbf{K}$) and the initial source ($\mathbf{S}$). The horizontal lower figure is the plot of the temporal focusing for $\mathbf{K}=\mathbf{S}$, and the vertical one is the spatial focusing at time $t=0$. In both cases, the position of the Time Reversal Window is $40 \mu s$ and its duration is $4 \mu$ s. (a) is obtained when the TR channel ($\mathbf{R}$) is far away from the initial source ($\mathbf{S})$ and (b) when they are at the same position ($\mathbf{R}=\mathbf{S}$). The amplitudes are normalized such that the focusing amplitude in (a) equals 1.}
\end{figure}

\begin{figure}
        \begin{center}
                \includegraphics{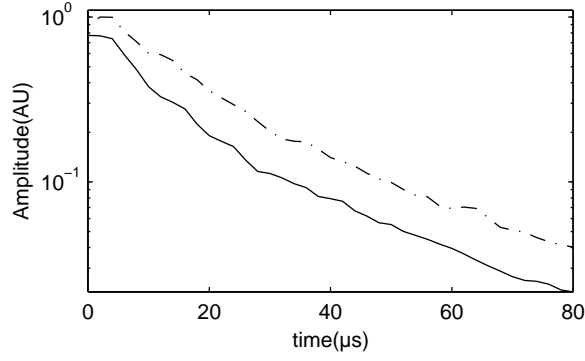}
        \end{center}
        \caption{\label{fig:mean_time_of_flight}Amplitude of the average TR field (recorded at time t=0 and at the focal position, i.e., $\mathbf{S}=\mathbf{K}$) versus the position of the TR window ($t_0$). Continuous line (respect. dotted line) corresponds to the case where the TR channel is far from the source whereas the dotted line corresponds to the situation where the initial source point and the TR channel are identical.}
\end{figure}

\begin{figure}
        \begin{center}
                \includegraphics{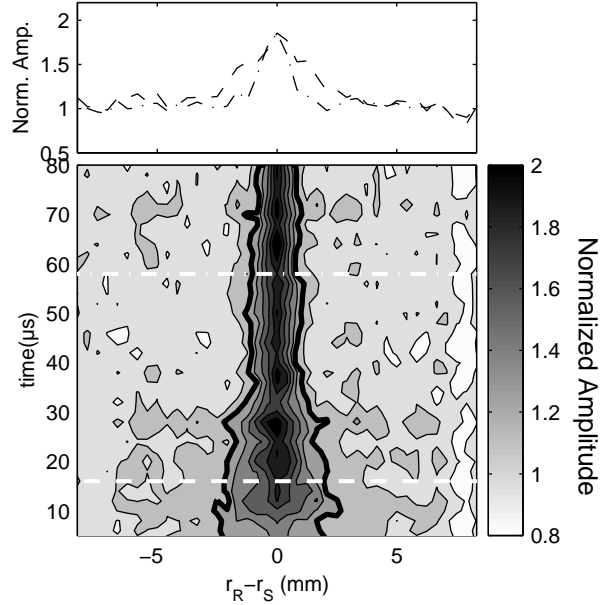}
        \end{center}
        \caption{\label{fig:mean_cone}Average TR focusing peak amplitude at the source ($\mathbf{S}=\mathbf{K}$) versus the distance between the source and the TR channel ($\mathbf{r}_{\mathbf{R}} - \mathbf{r}_{\mathbf{S}}$) versus the position $t_0$ of the TR window. At each time, the field is normalized such that the field equals 1 when $\mathbf{S}$ and $\mathbf{K}$ are far apart. The top plot shows two snapshots of the bottom map at times $t_0=17\mu s$ (dashed line) and $t_0=57\mu s$ (dashed dotted line)}
\end{figure}

As in the preliminary part, the source ($\mathbf{S}$) sends a 1-$\mu s$ pulse at $3.5MHz$ central frequency. The backscattered ultrasonic field is recorded at point $\mathbf{R}$. A short time reversal window centered at $t_0=40 \mu s$ and $\Delta T= 4 \mu s$ long is selected, time-reversed and transmitted back towards the medium. This experiment is repeated for one hundred realizations of disorder.  
We have recorded the average TR field for two extreme situations: either the one-channel TR device and the initial source are far away from each other (Fig.~\ref{fig:mean_spatio_tempo_1et2}a), or they are at the same location (Fig.~\ref{fig:mean_spatio_tempo_1et2}b). First we observe that the hyper-focusing property of TR through complex media is achieved: the focal spot is very thin ($2mm$ i.e., 4 wavelengths) whereas no spatial focusing would occur in free space with one single TR channel. 
\begin{figure}
        \begin{center}
         \includegraphics[width=8cm]{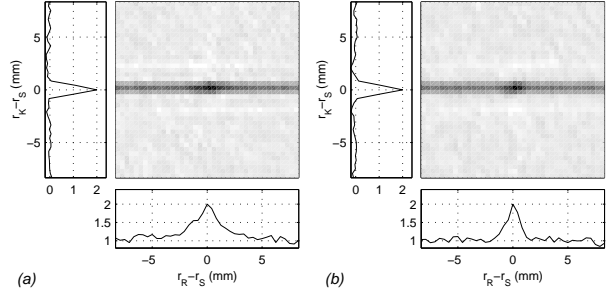}
        \end{center}
        \caption{Gray level representations of the average TR field at time $t=0$ versus the distance between the initial source and the measurement point  ($ \left\| \mathbf{S}-\mathbf{K} \right\|$ on the vertical scale) and the distance between the initial source and the TR channel ($\left\| \mathbf{S}-\mathbf{R} \right\|$ on the horizontal scale). (a) and (b) correspond to a 4 $\mu s$ long time reversal window centered at time $t_0=20 \mu s$ and $t_0=40 \mu s$ respectively.}
        \label{fig:mean_spatio_spatio}
\end{figure}
Furthermore we notice that the amplitude of the pulse is twice as large as when the TR channel is at the same position as the initial source ($\mathbf{R}=\mathbf{S}$). This is a manifestation of the coherent backscattering effect. As predicted by our theoretical study (see Fig.~\ref{fig:simul_ave_spatio_spatio}), we observe in Fig.~\ref{fig:mean_spatio_spatio} that the TR focusing amplitude versus the distance between $\mathbf{S}$ and $\mathbf{R}$ is peaked like the CBS. The width of the peak narrows as $t_0$ increases. 
In Fig.~\ref{fig:mean_time_of_flight} is plotted the dependence of the peak amplitude with respect to the TR window position when $\mathbf{S}=\mathbf{R}$ and when $\mathbf{S} \neq \mathbf{R}$ (Fig.~\ref{fig:mean_time_of_flight}). We observe the behavior predicted by the theory (see Fig.~\ref{fig:simul_tof}).  Especially we observe that the factor 2 is measured whatever the Time Reversal Window as soon as there are at least two scattering events.
Finally Fig.~\ref{fig:mean_cone} presents a map of the peak amplitude with respect to the postion of the TR window ($t_0$) and the distance between the source ($\mathbf{S}$) and the TR channel ($\mathbf{R}$).
Its width decreases as $1/\sqrt{D t_0}$ as $t_0$ increases, like the dynamic CBS\cite{Tourin97}.
\subsection{Variance}
\label{exp:var}
\begin{figure}
        \begin{center}
                \includegraphics{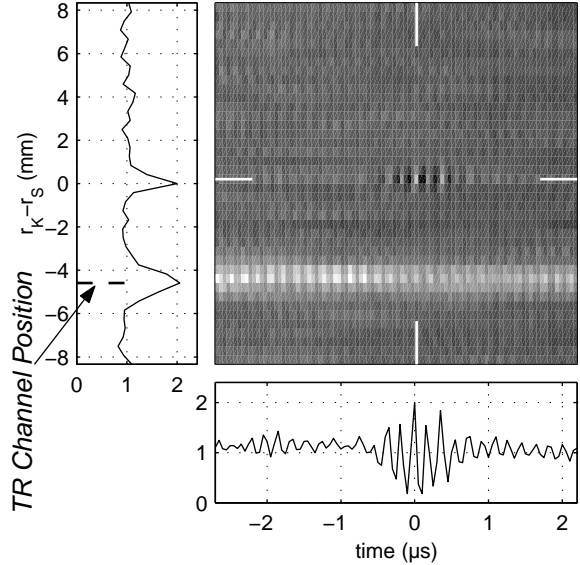}
        \end{center}
        \caption{\label{fig:var_space_time}Gray level representation of the variance of the TR field versus time $t$ (horizontal scale) and versus the distance between the measurement point $\mathbf{K}$ and the initial source $\mathbf{S}$ (vertical scale). The TR channel is located at $-4.6\:mm$ from the source. The horizontal plot is the time evolution of the variance at the focus ($\mathbf{S}=\mathbf{K}$). The vertical plot is the spatial dependence of the field at $t=0$. The TR window is $\Delta T = 2 \mu s$ long and centered at time $t_0=40 \mu s$.}
\end{figure}

\begin{figure}
        \begin{center}
                \includegraphics[width=8cm]{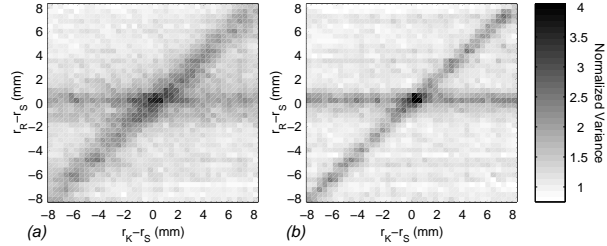}
        \end{center}
        \caption{\label{fig:var_spatial_spatial_bruit}Variance of the TR field outside the focusing time. The vertical axis corresponds to the distance between the TR channel ($\mathbf{R}$) and the source ($\mathbf{S}$). The horizontal axis is the distance between the measurement point ($\mathbf{K}$) and the source ($\mathbf{S}$). The $2 \mu s$ long TR windows are centered around time $20 \mu s$ (a) and $40 \mu s$ (b) respectively.}
\end{figure}

\begin{figure}
        \begin{center}
                \includegraphics[width=8cm]{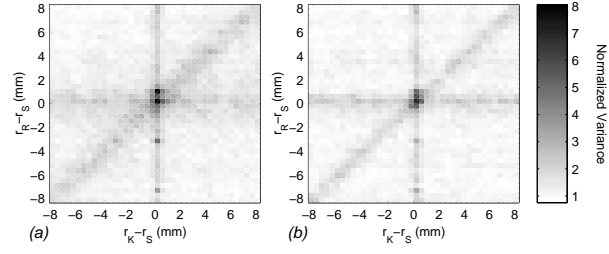}
        \end{center}
        \caption{\label{fig:var_spatial_spatial_pic}Variance of the TR field at the focusing time. The vertical axis corresponds to the distance between the TR channel $\mathbf{R}$ and the source $\mathbf{S}$. The horizontal axis is the distance between the measurement point $\mathbf{K}$ and the source $\mathbf{S}$. The TR windows are centered around time $20 \mu s$ (a) and $40 \mu s$ (b) respectively.}
\end{figure}

Figure~\ref{fig:mean_spatio_tempo_1et2} shows a map of the average TR field when the source point $\mathbf{S}$ and the single TR channel $\mathbf{R}$ are far away from each other. The corresponding map of the variance is shown in Fig.~\ref{fig:var_space_time}. 
As predicted by the theory, two contributions add up to a flat variance that was normalized here to 1. Upon varying the position of the receiver $\mathbf{K}$, the first contribution appears around the position of the TR channel $\mathbf{R}$. This contribution is seen to be roughly constant in time. The second contribution only occurs at the source position $\mathbf{S}$ and for times around the focusing time $t=0$. As was expected, this contribution oscillates twice as fast as the central frequency of the initial pulse. In both case, the maximum enhancement factor is 2. 
Finally the experimental results shown in Figs.~\ref{fig:var_spatial_spatial_bruit} and ~\ref{fig:var_spatial_spatial_pic} confirm the theoretical result that was plotted earlier in Fig.~\ref{fig:simul_var3}.
\section{Conclusions}
One-channel time-reversal experiments have been revisited within the framework of the multiple scattering theory. On the one hand, previously established results such as hyper-resolution and self-averaging have been rigorously demonstrated within this theoretical framework. On the other hand, the link between one-channel TR and CBS has been established. Especially, we have shown that when the initial source and the time reversal point coincide, the time-reversed amplitude is twice as large. Surprisingly, this enhancement is due to the ladder diagrams and not to the most-crossed ones, contrary to CBS. These theoretical predictions have been confirmed by experimental results which have been obtained with ultrasonic waves propagating through a random 2D collection of parallel steel rods. The generalization to multiple-channel TR and to random media subject to spatial long-range correlations will be the object of further studies.
\begin{acknowledgments}
The authors wish to acknowledge David Lacoste for fruitful discussions as well as the GDR IMCODE of CNRS for its financial support.\end{acknowledgments}
\clearpage 
\appendix
\section{Mean value of $\chi$ and $\chi_2$}
\label{appendix:calculs}
Let $A$ and $B$ be two time-dependent, real-valued, random functions that in the text correspond to ${\T}(\mathbf{r}_{1} \rightarrow \mathbf{r}_{2})$ and $ {\T}(\mathbf{r}_{3} \rightarrow \mathbf{r}_{4})$ respectively. We want to find the expression for the correlation between these two functions from their frequency correlations. From the definition of the Fourier transform, the time correlation is
\begin{eqnarray}
        \left\langle A(t)  B(t')\right\rangle & = & \int \left\langle  A(\omega)  B^{*}(\omega')\right\rangle e^{i ( - \omega t + \omega' t')}  d\omega d\omega' \nonumber\\
                         & = & \int  C_{A,B}(\omega -\omega') e^{i t (- \omega + \omega')} e^{i \omega' (t'-t)}  d\omega d\omega'\nonumber\\
                        & = &  C_{A,B}(t) \delta(t'-t) 
                        \label{eq:annex_corre}
\end{eqnarray}
Here and later, the integrations are implicitly performed from $-\infty$ to $\infty$. We have assumed that the correlation function in the frequency domain $C_{A,B}$ depends only on the difference $\delta \omega = \omega-\omega'$. $C_{A,B}(t)$ is the inverse Fourier transform of $C_{A,B}(\delta \omega)$.
To calculate the mean TR field, we need to compute
\begin{eqnarray}
        \left\langle \chi(t) \right\rangle & = & \int \left\langle B(\tau) A(\tau-t)\right\rangle  W(\frac{\tau-t-t_0}{\Delta T} ) d\tau 
\end{eqnarray}
Using Eq.~(\ref{eq:annex_corre}), the last expression reads 
\begin{eqnarray}
        \left\langle \chi(t) \right\rangle & = & \left[ \int  C_{A,B}(\tau) W(\frac{\tau-t_0}{\Delta T} ) d\tau \right] \delta(t)
\end{eqnarray}
Evaluation of the variance of the TR process implies to work out: 
\begin{eqnarray}
                  \left\langle \chi_2(\tau,\tau') \right\rangle    =   \int \left\langle B(\tau_1) A(\tau_1-\tau)  B'(\tau_1') A'(\tau_1'-\tau') \right\rangle \\ \nonumber 
                 \times W(\frac{\tau_1-\tau-t_0}{\Delta T} )  W(\frac{\tau_1'-\tau'-t_0}{\Delta T} ) \ d\tau_1' d\tau_1
\end{eqnarray}
Assuming $A$,$B$,$A'$,$B'$ are jointly Gaussian variables, this becomes 
\begin{multline}
        \left\langle B(\tau_1) A(\tau_1-\tau)  B'(\tau_1') A'(\tau_1'-\tau') \right\rangle = \\
        {\left\langle B(\tau_1) A(\tau_1-\tau)   \right\rangle\left\langle  B'(\tau_1') A'(\tau_1'-\tau') \right\rangle} + \\
                \underbrace{\left\langle B(\tau_1) B'(\tau_1')   \right\rangle\left\langle  A(\tau_1-\tau)  A'(\tau_1'-\tau') \right\rangle}_{C_{B,B'}(\tau_1) C_{A,A'}(\tau_1-\tau) \delta(\tau_1-\tau_1') \delta(\tau_1-\tau_1'-\tau+\tau')} + \\
        \underbrace{\left\langle B(\tau_1) A'(\tau_1'-\tau')  \right\rangle \left\langle  B'(\tau_1') A(\tau_1-\tau) \right\rangle }_{C_{B,A'}(\tau_1) C_{B',A}(\tau_1-\tau) \delta(\tau_1-\tau_1'+\tau') \delta(\tau_1'-\tau_1+\tau)}
\end{multline}
Hence, $\left\langle \chi_2 \right\rangle$ is the sum of tree terms:
\begin{eqnarray}
                 \left\langle \chi_2(\tau,\tau') \right\rangle   & = &  \mathbf{X}_1(\tau,\tau') +  \mathbf{X}_2(\tau,\tau')+ \mathbf{X}_3(\tau,\tau') 
\end{eqnarray}
where
\begin{eqnarray}
         \mathbf{X}_1(\tau,\tau') &= & \left\langle \chi( \tau) \right\rangle \left\langle \chi(\tau') \right\rangle \\
         \mathbf{X}_2(\tau,\tau')&= &{\left( C_{B,B'} (\tau) \convo  \left[ C_{A,A'}(-\tau)  W^2(\frac{-\tau-t_0}{\Delta T} )  \right]\right)}   \delta(-\tau+\tau') \\
 \mathbf{X}_3(\tau,\tau')&= &{\left[ C_{B,A'}(\tau) W(\frac{\tau-t_0}{\Delta T} ) \right] \convo \left[ C_{B',A}(-\tau)  W(\frac{-\tau-t_0}{\Delta T} ) \right]}   \delta(\tau'+\tau)
\end{eqnarray}
These expressions can be greatly simplified if we assume that the correlation function is constant inside the duration of the Time Reversal Window $\Delta T$
\begin{eqnarray}
         \mathbf{X}_2(\tau,\tau') & \approx & \Delta T C_{A,A'}(t_0)C_{B,B'}(\tau+t_0) \delta(-\tau+\tau') \\
   \mathbf{X}_3(\tau,\tau') & \approx & \Lambda(\frac{\tau}{\Delta T})  C_{B,A'}(t_0) C_{B',A}(t_0) \delta(\tau'+\tau) 
\end{eqnarray}
where $\Lambda$ is the triangle function, i.e., $ \Lambda(x)=1-\left| x \right|$ if $ \left|x\right|<1 $ and $ 0 $ if $\left|x\right|>1$.
\section{Far Field Approximation}
\label{sec:far_field_approx}
The far field approximation leads to write the convolution of two effective Green's functions as 
\begin{eqnarray}
G_{e}(\mathbf{X}_1 \rightarrow \mathbf{r}_{1} ; -t) \convo  G_{e}(\mathbf{X}_2  \rightarrow \mathbf{r}_{1} ; t) = \rho \:
\delta (t- \frac{x_1 (\mathbf{X}_2-\mathbf{X}_1) }{c a} ) \times 
\left\{ 
\begin{array}{l} 
 e^{-\frac{z_1}{\textstyle \ell \mu_1}} \:\mbox{when} \:z>0 
 \cr
 1 \:\mbox{when}\: z<0
\end{array}
\right.
\end{eqnarray}
where $a$ is the distance between the plane that contains $\mathbf{X}_1$, $z_1$ and $x_1$ which are respectively the normal and parallel components of the position vector with respect to the plane interface of the multiple scattering medium. $\rho$ is a constant equal to $1/a^4$ in 3D. $\ell$ is the elastic mean free path which is assumed to be independent of frequency (resonances are ignored). $\mu_1= cos(\theta_1)$ where $\tan(\theta_1) = x_1  / a  $. Then Eq.~(\ref{eq:def_zeta}) can be rewritten as
\begin{eqnarray}
\xi_{t_0}(\mathbf{X}_1,\mathbf{X}_2,\mathbf{X}_3,\mathbf{X}_4;t) = \nonumber \\ 
 \underbrace{\int \delta( t +   \mathbf{x}_{1} \frac{\mathbf{X}_4 - \mathbf{X}_1}{c a}  + \mathbf{x}_{2} \frac{\mathbf{X}_3 - \mathbf{X}_2}{c a} ) F'(\mathbf{x}_{1},\mathbf{x}_{2};t_0)  dx_1 dx_2   \convo f(-t)}_{C} \nonumber \\
+  \underbrace{\int \delta( t +   \mathbf{x}_{1} \frac{\mathbf{X}_3 - \mathbf{X}_1}{c a}  + \mathbf{x}_{2} \frac{\mathbf{X}_4 - \mathbf{X}_2}{c a} ) F'(\mathbf{x}_{1},\mathbf{x}_{2};t_0)  dx_1 dx_2 \convo f(-t) }_{R} \nonumber    
\end{eqnarray}
where $F'$ is given by:
\begin{eqnarray}
F'(x_1,x_2;t_0)= \rho \int e^{-z_2 \ell / \mu_2} e^{-z_1 \ell / \mu_1} F(x_1,x_2,z_1,z_2 ; t_0) dz_1 dz_2
\end{eqnarray}
The integrations are always implicitly performed from $-\infty$ to $\infty$. A narrow band pulse can be treated as almost monochromatic. Then $\xi$ can simply be expressed in terms of the spatial Fourier transform of $F'$: 
\begin{eqnarray}
\xi_{t_0}(\mathbf{X}_1,\mathbf{X}_2,\mathbf{X}_3,\mathbf{X}_4;t)  = [ \underbrace{\tilde{F'}(\mathbf{X}_4 - \mathbf{X}_1,\mathbf{X}_3 - \mathbf{X}_2; t_0)}_{C} \nonumber \\ +  \underbrace{\tilde{F'}(\mathbf{X}_3 - \mathbf{X}_1,\mathbf{X}_4 - \mathbf{X}_2; t_0)}_{R}  ] f(-t)
\label{eq:zetaFT}
\end{eqnarray}
where the 2-D Fourier transform is:
\begin{eqnarray}
         \tilde{F'}(X,Y) = \int e^{- i {\frac{k  \mathbf{x}_{1}}{a}} X}  F'(x_1,x_2 ; t_0) e^{- i {\frac{k  \mathbf{x}_{2}}{a}} Y} dx_1 dx_2
\end{eqnarray}%
\bibliography{derosny-text} 
\bibliographystyle{unsrt} 
\end{document}